%% file: aaai25.tex
\documentclass[letterpaper]{article} 
\usepackage{aaai25}  
\usepackage{times}  
\usepackage{helvet}  
\usepackage{courier}  
\usepackage[hyphens]{url}  
\usepackage{graphicx} 
\usepackage{natbib}  
\usepackage{caption} 
\usepackage{algorithm}
\usepackage{algorithmic}
\usepackage{xcolor}
\usepackage{amsmath}
\usepackage{graphicx}
\usepackage{amssymb}
\usepackage{newfloat}
\usepackage{listings}
\DeclareCaptionStyle{ruled}{labelfont=normalfont,labelsep=colon,strut=off} 
\floatstyle{ruled}
\newfloat{listing}{tb}{lst}{}
\floatname{listing}{Listing}
\title{DP-Adapter: Dual-Pathway Adapter for Boosting Fidelity and Text Consistency in Customizable Human Image Generation}

\author{
    Ye Wang\textsuperscript{\rm 1},
    Xuping Xie\textsuperscript{\rm 2},
    Lanjun Wang \textsuperscript{\rm 3},
    Zili Yi\textsuperscript{\rm 4}\footnote{Corresponding authors},
    Rui Ma\textsuperscript{\rm 1,5}\footnotemark[1]
}

\affiliations{
    \textsuperscript{\rm 1} School of Artificial Intelligence, Jilin University\\
    \textsuperscript{\rm 2} College of Computer Science and Technology, Jilin University\\ 
    \textsuperscript{\rm 3} School of New Media and Communication, Tianjin University. \\
    \textsuperscript{\rm 4} School of Intelligence Science and Technology, Nanjing University,\\
    \textsuperscript{\rm 5}  Engineering Research Center of Knowledge-Driven Human-Machine Intelligence, MOE, China\\
    \{yewang22, xiexp21\}@mails.jlu.edu.cn, wanglanjun@tju.edu.cn, yi@nju.edu.cn,  ruim@jlu.edu.cn
    \\
}

\usepackage{bibentry}
\usepackage{amsmath}

\begin{document}

\maketitle


\input{0_abstract}

\input{1_1_introduction}

\input{2_related_work}

\input{3_method}

\input{4_experiments}

\input{5_conclusion}

\bibliography{aaai25}
\end{document}

%% file: 0_abstract.tex
\begin{abstract}

With the growing popularity of personalized human content creation and sharing, there is a rising demand for advanced techniques in customized human image generation. However, current methods struggle to simultaneously maintain the fidelity of human identity and ensure the consistency of textual prompts, often resulting in suboptimal outcomes. This shortcoming is primarily due to the lack of effective constraints during the simultaneous integration of visual and textual prompts, leading to unhealthy mutual interference that compromises the full expression of both types of input. Building on prior research that suggests visual and textual conditions influence different regions of an image in distinct ways, we introduce a novel Dual-Pathway Adapter (DP-Adapter) to enhance both high-fidelity identity preservation and textual consistency in personalized human image generation. Our approach begins by decoupling the target human image into visually sensitive and text-sensitive regions. For visually sensitive regions, DP-Adapter employs an Identity-Enhancing Adapter (IEA) to preserve detailed identity features. For text-sensitive regions, we introduce a Textual-Consistency Adapter (TCA) to minimize visual interference and ensure the consistency of textual semantics. To seamlessly integrate these pathways, we develop a Fine-Grained Feature-Level Blending (FFB) module that efficiently combines hierarchical semantic features from both pathways, resulting in more natural and coherent synthesis outcomes. Additionally, DP-Adapter supports various innovative applications, including controllable headshot-to-full-body portrait generation, age editing, old-photo to reality, and expression editing.
Extensive experiments demonstrate that DP-Adapter outperforms state-of-the-art methods in both visual fidelity and text consistency, highlighting its effectiveness and versatility in the field of human image generation.
\end{abstract}


%% file: 1_1_introduction.tex
\section{Introduction}

\begin{figure}[t]
    \centering
    \includegraphics[width=\linewidth]{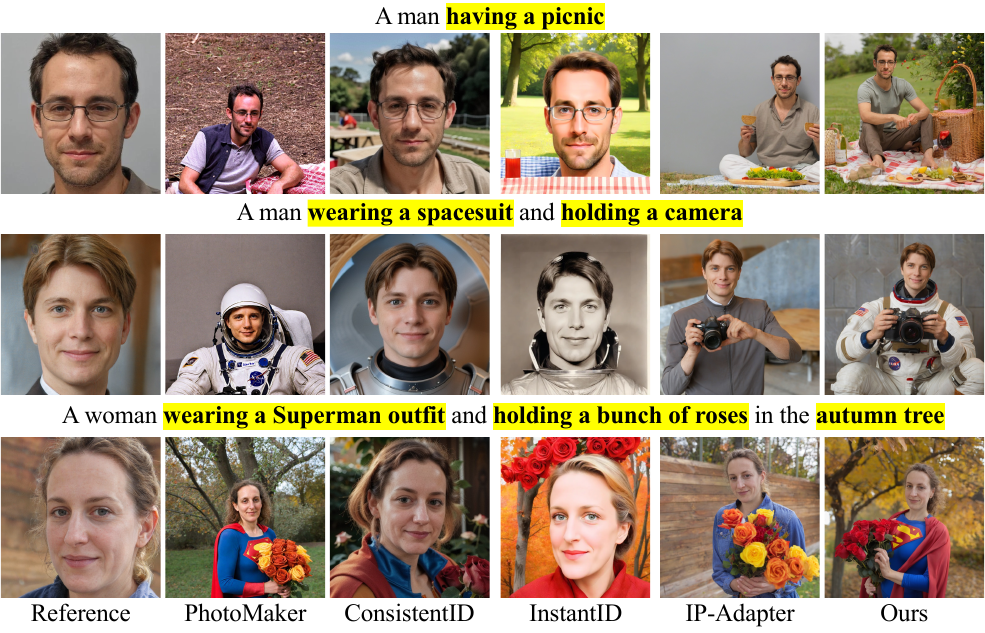}
    \caption{
Compared to other methods, our approach excels in preserving human identities while ensuring textual semantic consistency.} 
    \label{fig:teaser} 
\end{figure}

In today's digital age, the creation and sharing of personal content have become ubiquitous. AI-powered methods for generating human-centric content provide individuals with innovative avenues for self-expression and presentation. Customized human image generation \cite{li2024photomaker, xiao2023fastcomposer, wang2024high, wang2024instantid, huang2024consistentid,wu2024infinite,guo2024pulid}—which focuses on creating entirely new images of individuals based on reference images and textual prompts—has garnered significant scholarly interest. This technology offers diverse applications, including virtual try-on, advertising design, and artistic creation, making it a valuable tool in various creative and commercial fields.

Current methods for customized human image generation can be categorized into two types, based on whether fine-tuning is required during testing. The first type involves methods that fine-tune pre-trained text-to-image (T2I) models, such as Stable Diffusion \cite{rombach2022high}, using multiple images of the same identity (ID). Notable examples include Dreambooth \cite{ruiz2023dreambooth}, Textual Inversion \cite{gal2022image}, and Low-Rank Adaptation (LoRA) \cite{hu2021lora}. While these approaches are effective in achieving high fidelity, the fine-tuning process is resource-intensive and time-consuming, which limits their practicality. Additionally, the need for multiple images per ID poses challenges in scenarios where data is limited.

In a parallel line of research, tuning-free methods typically utilize a pre-trained image encoder to extract reference image embeddings as visual guidance, enabling image generation in a single forward pass. Prominent examples include FastComposer \cite{xiao2023fastcomposer}, PhotoMaker \cite{li2024photomaker}, Face-Diffuser \cite{wang2024high}, ConsistentID \cite{huang2024consistentid}, InstantID \cite{wang2024instantid}, and IP-Adapter \cite{ye2023ip}. Although these methods are computationally efficient, they often struggle to balance high-fidelity image generation with maintaining strong text consistency. As illustrated in Figure \ref{fig:teaser}, methods such as ConsistentID \cite{huang2024consistentid}, InstantID \cite{wang2024instantid}, and IP-Adapter \cite{ye2023ip} can produce high-fidelity images of individuals, but frequently lose key textual elements, such as Superman outfits or autumn trees, leading to reduced semantic consistency with text prompts. On the other hand, PhotoMaker \cite{li2024photomaker} demonstrates a better understanding of textual semantics, but tends to generate images with lower fidelity, such as altering the reference person's hairstyle (see Figure \ref{fig:teaser}, third row).

The primary reason these methods struggle to overcome the aforementioned challenges is the lack of effective constraints during the joint integration of visual and textual conditions. Traditionally, these approaches either directly blend visual features with corresponding text token features \cite{xiao2023fastcomposer,li2024photomaker}, or use a trainable module to process visual features separately \cite{ye2023ip,huang2024consistentid}. However, both techniques fail to impose sufficient supervision and constraints upon the interaction between text and visual information. This lack of constraints allows visual information to encroach upon areas where textual information should dominate, and vice versa, leading to severe harmful interference between the two modalities, ultimately resulting in the suboptimal expression of both.

Recent study, Face-Diffuser \cite{wang2024high}, suggests that visual and textual conditions influence the generation of different image regions in distinct ways. Specifically, two distinct types of region can be identified: visually sensitive regions and text-sensitive regions. Visually sensitive regions, such as facial areas, are primarily influenced by visual conditions, which are crucial to accurately depicting detailed features such as face shape and skin tone. In contrast, text-sensitive regions, such as the background, actions, and poses, are predominantly guided by textual conditions, with visual conditions serving a supplementary role in these areas.  To leverage this property, Face-Diffuser \cite{wang2024high} proposes to tune two independent models for subject and scene generation; however, it sacrifices training efficiency and is prone to introducing unnatural qualities or even irrationality during regional fusion.

Unlike Face-Diffuser \cite{wang2024high}, we propose a Dual-Pathway image prompt Adapter  (DP-Adapter) module to enhance text consistency and improve image fidelity. For visually sensitive regions, such as facial features, we implement an Identity-Enhancing Adapter (IEA) that leverages the original visual conditions to ensure a detailed and accurate representation. For text-sensitive regions, including backgrounds and poses, we use a Textual-Consistency Adapter (TCA) to minimize the impact of visual conditions, thereby reducing visual intrusion and preserving textual semantics.

After processing the visual and textual prompts separately, we merge the information from both conditions using a Fine-grained Feature-level Blending (FFB) module. Unlike most methods \cite{avrahami2022blended, zhao2023magicfusion, wang2024high}, which rely on direct blending in the noise space and are prone to introducing visual artifacts, our FFB module combines hierarchical semantic features from both pathways. This approach ensures a smooth and artifact-free integration of visual and textual conditions, resulting in more coherent and high-quality generated images.

Based on both quantitative and qualitative results, our approach not only achieves high-fidelity human image personalization but also excels at maintaining text consistency compared to current state-of-the-art (SOTA) methods. Our key contributions can be summarized as follows:
\begin{itemize}
\item We introduce a novel dual-pathway image prompt adapter for human image customization, featuring an Identity-Enhancing Adapter (IEA) to emphasize the processing of visual prompts for visually sensitive regions, and a Textual-Consistency Adapter (TCA) to prioritize text prompts for text-sensitive regions. This approach effectively mitigates the harmful mutual interference between visual and textual conditions seen in existing methods, leading to significant improvements in both human identity preservation and text consistency.

\item We present a fine-grained feature-level blending module that hierarchically combines visual and semantic information across multiple feature levels. This approach greatly reduces artifacts in the generated images, resulting in more natural and coherent synthesis outcomes.

\item Extensive experiments across various application scenarios demonstrate that our method surpasses other state-of-the-art approaches, delivering superior performance and robust generation capabilities.
\end{itemize}

%% file: 2_related_work.tex
\section{Related Work}

\subsection{Personilized Text-to-Image Generation} Diffusion models \cite{rombach2022high,nichol2021glide,podell2023sdxl,peebles2023scalable} have attracted widespread attention from both industry and academia due to their outstanding performance and high fidelity. This progress has spurred rapid development in customized image generation techniques. Currently, mainstream customized image generation methods can be categorized into two types. The first type relies on fine-tuning during test-time, with representative works including Dreambooth \cite{ruiz2023dreambooth}, Textual Inversion \cite{gal2022image}, Custom Diffusion \cite{kumari2023multi}, and LoRA \cite{hu2021lora}. Despite their advancements, these methods often require collecting multiple images to ensure learning performance and necessitate individual fine-tuning for each example, which is time-consuming and labor-intensive, thus limiting their practicality. The second type employs tuning-free techniques, skipping the additional fine-tuning or inversion process. Representative works include IP-Adapter \cite{ye2023ip}, FastComposer \cite{xiao2023fastcomposer}, PhotoMaker \cite{li2024photomaker}, Face-Diffuser \cite{wang2024high}, InstantID \cite{wang2024instantid}, ConsistentID  \cite{huang2024consistentid}, PuLID \cite{guo2024pulid} and Infinite-ID \cite{wu2024infinite}. This type of method achieves the customized generation using only an image in a single forward process. Even with improved computational efficiency, these methods often struggle to achieve both high-fidelity generation and high text consistency simultaneously. They either offer excellent text consistency but at the cost of fidelity \cite{li2024photomaker}, or high fidelity but with reduced text consistency \cite{huang2024consistentid,wang2024instantid,ye2023ip}. In contrast, our approach excels in both metrics by using a unique dual-pathway processing and collaboration mechanism.

\subsection{Adapter for Diffusion Models}
Originated from Natural Language Processing \cite{houlsby2019parameter}, adapter technology has also been applied to diffusion models. ControlNet \cite{zhang2023adding} and T2I-Adapter \cite{mou2024t2i} pioneered the use of adapters to integrate more spatial signals for controllable generation. IP-Adapter \cite{ye2023ip} introduces a decoupled, trainable cross-attention module that receives image prompts, enabling the generation of images similar to the input images. 
In this paper, we propose DP-Adapter, a novel dual-pathway image adapter. Unlike existing methods, each adapter in DP-Adapter has a distinct learning objective: one aims to enhance text consistency, while the other focuses on maintaining high-fidelity identity preservation. By employing a fine-grained feature-level blending module, we achieve an organic integration of these two adapter, resulting in image generation that is both high-fidelity and textually consistent.

\subsection{Diffusion Blending}

Diffusion Blending is widely used in the field of image local editing. Most current methods \cite{lugmayr2022repaint, avrahami2023blended, avrahami2022blended,couairon2022diffedit,zhao2023magicfusion} perform blending operations directly in the noisy image latents space. However, empirical experiments have shown that this approach may introduce artifacts and inconsistencies into the results. This is because the information in the intermediate noisy images lacks the necessary semantics to achieve consistent and seamless fusion. Therefore, we propose a fine-grained feature-level blending (FFB) module, which combines hierarchical semantic features from the two pathways, resulting in more natural and coherent synthesis results.

%% file: 3_method.tex
\begin{figure*}[t]
    \centering
    \includegraphics[width=\textwidth]{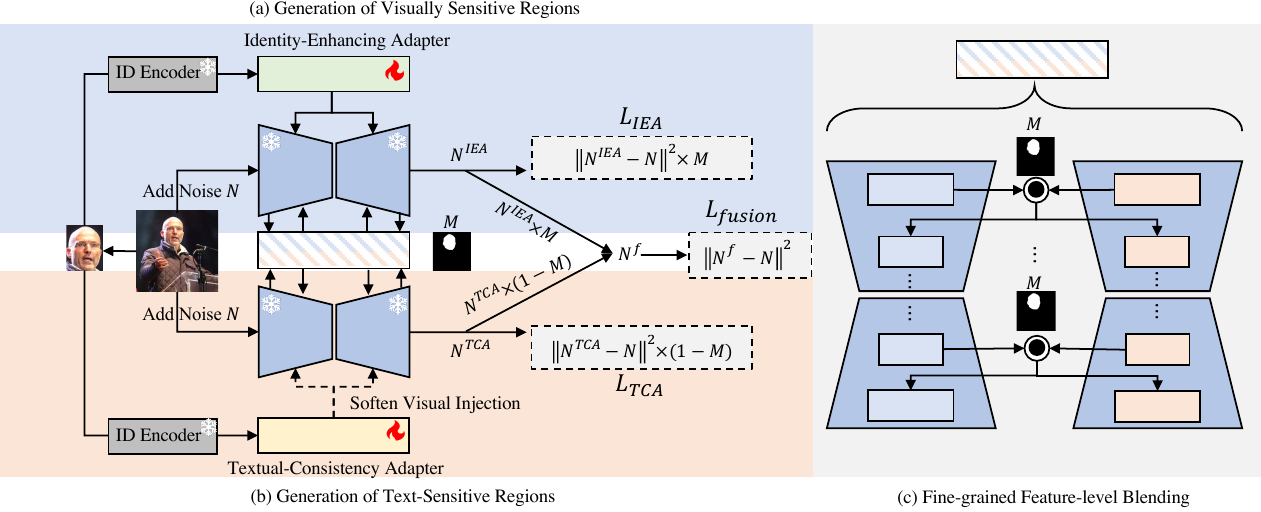}
    \caption{The DP-Adapter framework utilizes a region-separate processing approach, dividing the target image generation into visually sensitive and text-sensitive regions. For visually sensitive regions, an identity-enhancing adapter is employed to ensure high-fidelity face generation. For text-sensitive regions, a textual-consistency adapter is used to maintain semantic accuracy. To seamlessly blend these regions, we introduce a fine-grained feature-level blending module, which effectively integrates the different areas and minimizes artifacts.
    For simplicity, the textual injection branch is not displayed.}
    \label{fig:method_framework}
\end{figure*}

\section{Method}

In this section, we first introduce the preliminaries of Stable Diffusion \cite{rombach2022high} and image adapter techniques. We then introduce our proposed dual-pathway image adapter model. 
\subsection{Preliminary}

\noindent{\textbf{Stable Diffusion}}.  As a state-of-the-art text-to-image generation model, it begins by encoding an input image $x$ into a latent representation $z$ using a VAE encoder. Noise $N$ is then introduced at time step $t$ to create a noisy latent $z_t$. To guide the generation process with text conditions, Stable Diffusion \cite{rombach2022high} incorporates a CLIP text encoder $\tau$ to encode textual prompts $C$, which are integrated into the cross-attention layers for interaction with the noisy latents. Finally, a conditional U-Net backbone $\epsilon_\theta$ is trained to predict the noise. The training objective is as follows:
\begin{equation}
    L_{SD}(\theta) := \mathbb{E}_{t,x_0,\epsilon} \left[ \lVert N - \epsilon_\theta(z_t, t, C) \rVert^2 \right] \label{eq:LLDM}
\end{equation}


\noindent{\textbf{Image Adapter}}. 
Image adapter techniques, as presented in \citet{ye2023ip, song2024moma, mou2024t2i}, leverage lightweight adapter networks to process visual condition inputs. These techniques enhance pre-trained text-to-image diffusion models by incorporating image prompt capabilities. They typically employ a trainable cross-attention module that processes image conditions and merge the output with the text cross-attention branch, as detailed below:
\begin{equation}
\operatorname{Attn}_{final} = \operatorname{Attn}_{text} + \alpha \cdot \operatorname{Attn}_{image}
\end{equation}
where $\alpha$ is a weight that controls the intensity of image condition injection.

\subsection{DP-Adapter Framework}

The pipeline of DP-Adapter includes three parts: (a) Generation of visually sensitive regions
(b) Generation of text-sensitive regions, and (c) fine-grained feature-level blending.  In the following sections, we will provide a detailed exposition of the functions and characteristics of these important components.

\noindent{\textbf{Generation of Visually Sensitive Regions}}. As shown in the upper branch of Figure \ref{fig:method_framework}.a, the objective is to achieve high-fidelity human generation, with a specific focus on the facial region. Specifically, we crop the face image from the original image and utilize a pre-trained visual encoder to extract the input face embedding. Subsequently, a trainable MLP is employed to project the visual embedding. To accommodate the visual embedding, we construct a lightweight identity-enhancing adapter (IEA) comprising a trainable cross-attention module. Finally, the output of the IEA is added to that of the original text cross-attention, in the same manner as IP-Adapter \cite{ye2023ip}. To achieve precise control over the regions where visual conditions are applied, inspired by the principle of division of labor, we propose a region-based denoising loss function to replace the traditional global denoising. We focus on the denoising learning of Visually Sensitive Regions (i.e., the facial region). Such a specific learning objective allows for precise control over the influence of the IEA, thereby significantly improving the identity fidelity of the generated images. Specifically, we use a face region mask to impose constraints on the original denoising loss function, ensuring that the model optimizes only for the designated region during the training process. The formulation of loss function is defined as follows:
\begin{equation}
    L_{IEA}(\theta) = \lVert M * (N - N^{IEA})\rVert^2 
    \label{eq:IEA}
\end{equation}
where, $M$ is the face region mask, $N^{IEA} = \epsilon_\theta(z_t, t, C, I_f)$ is the predicted noise, $I_f$ is the cropped face image.

\noindent{\textbf{Generation of Text-Sensitive Regions}}. The goal of generating text-sensitive regions is to enhance the semantic consistency between the generated image and the corresponding text prompt. As shown in the lower branch of the Figure \ref{fig:method_framework}.b, we propose the textual-consistency adapter (TCA). While TCA shares the same structure as IEA, their responsibilities differ. TCA employs a softened visual injection method to process face embeddings rather than completely dropping the visual condition. This approach mitigates the intrusion and interference of visual signals on text-sensitive regions, thereby enhancing the expression of textual signals. Specifically, we employ a simple yet effective method to achieve visual signal softening by lowering the $\alpha$ coefficient during the merging of visual and text cross-attention outputs, as illustrated below:

\begin{align}
    \operatorname{Attn}_{final} &= \operatorname{Attn}_{text} + \alpha \cdot \operatorname{Attn}_{image}, \alpha = 0.5
     \label{eq:alpha}
\end{align}
this straightforward adjustment effectively reduces visual interference, significantly improving textual semantic consistency. Moreover, to minimize the interference of textual signals on visually sensitive regions, TCA also adopts a region-based denoising loss function. During training, we focus on denoising learning in Text-Sensitive Regions, i.e., non-face regions, to ensure the model accurately applies textual semantics to the correct positions. The formulation of this loss function is defined as follows:

\begin{equation}
    L_{TCA}(\theta) = \lVert (1-M) * (N - N^{TCA})\rVert^2 
    \label{eq:IEA}
\end{equation}

where $N^{TCA}$ is the noise predicted by this pathway.

\noindent{\textbf{Fine-grained Feature-level Blending Module}}.
To obtain the final prediction, we need to merge the regions generated by IEA and TCA. A straightforward approach is to merge and blend at the noise latents level based on the mask. However, this simple blending results in numerous artifacts in the generated images. This is because the noise latent space lacks sufficient semantics to achieve consistent fusion, leading to degraded generation results. Conversely, the deep features of diffusion models contain rich semantic information. Therefore, we propose the fine-grained feature-level blending module, as illustrated in Figure \ref{fig:method_framework}.c. Specifically, we fuse the output features of corresponding blocks from the two branches using a mask and input the fused features into the next block. Since the size of the feature maps varies between blocks, we downsample or upsample the binary mask to match the resolution of the features. We then perform weighted fusion of the feature maps from the two branches, using the adjusted mask to control their respective contributions. 
This fusion process ensures that the features from different regions are effectively combined, thereby maintaining consistent semantic information and generating more natural and coherent synthesis results.
Finally, we merge the noise predictions from the two pathways based on the mask to obtain the global prediction and calculate the overall denoising loss, defined as follows:
\begin{align}
    N^{f} = (1-M)*N^{TCA} + M * N^{IEA} \\
    L_{fusion}(\theta) = \lVert (N - N^{f})\rVert^2 
     \label{eq:alpha}
\end{align}

\noindent{\textbf{Loss Function}}. The loss function of DP-Adapter is defined as follows:
\begin{equation}
    Loss = L_{IEA} + L_{TCA} + L_{fusion}
    \label{eq:fusion}
\end{equation}
where $L_{IEA}$ is dedicated to facial areas, refining the fidelity of facial features during the generation. $L_{TCA}$ focuses on optimizing non-facial regions to enhance the consistency between the generated image and its textual description. $L_{fusion}$, which is applied in the final stage after feature blending, performs a global optimization of the predictions. This step ensures that the generated images exhibit a natural harmony and visual cohesion.

\begin{figure}[t]
    \centering
    \includegraphics[width=\linewidth]{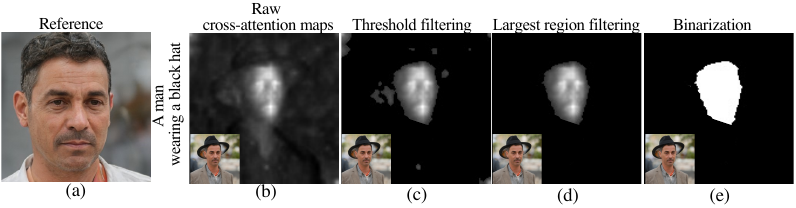} 
    \caption{The process of generating face mask images during the inference stage.} 
    \label{fig:inference_mask} 
\end{figure}

\noindent{\textbf{Inference of DP-Adapter}}. During the inference stage, it is common to lack masks as auxiliary information. Traditional methods \cite{gu2024mix,wang2024instantid} typically rely on manually defined bounding boxes (bbox), poses, and other spatial control information to guide image generation. While this provides some guidance, it limits the diversity of generated images and the flexibility of text prompts. To overcome this limitation, we propose a mask generation method based on layout priors, which automatically generates masks by leveraging internal model layout information. Specifically, our TCA possesses robust text semantic understanding capabilities, enabling it to generate image content that is highly consistent with the layout, background, and pose described in the text input. Based on this, we extract cross-attention maps corresponding to visual prompts from the TCA, as shown in Figure \ref{fig:inference_mask}.b. By applying a threshold, we filter out high-response areas, such as the face region, while excluding low-response areas, as demonstrated in Figure \ref{fig:inference_mask}.c. To further enhance mask quality, we introduce a largest region filtering technique. This technique identifies the largest contiguous area in the image and retains only the responses within this region, effectively excluding irrelevant responses outside of it, as shown in Figure \ref{fig:inference_mask}.d. Finally, through binarization, we obtain a region mask, providing more precise face position guidance for image generation (see Figure \ref{fig:inference_mask}.e). For further details, please refer to the supplementary material.

\begin{figure*}[t]
    \centering
    \includegraphics[width=\textwidth]{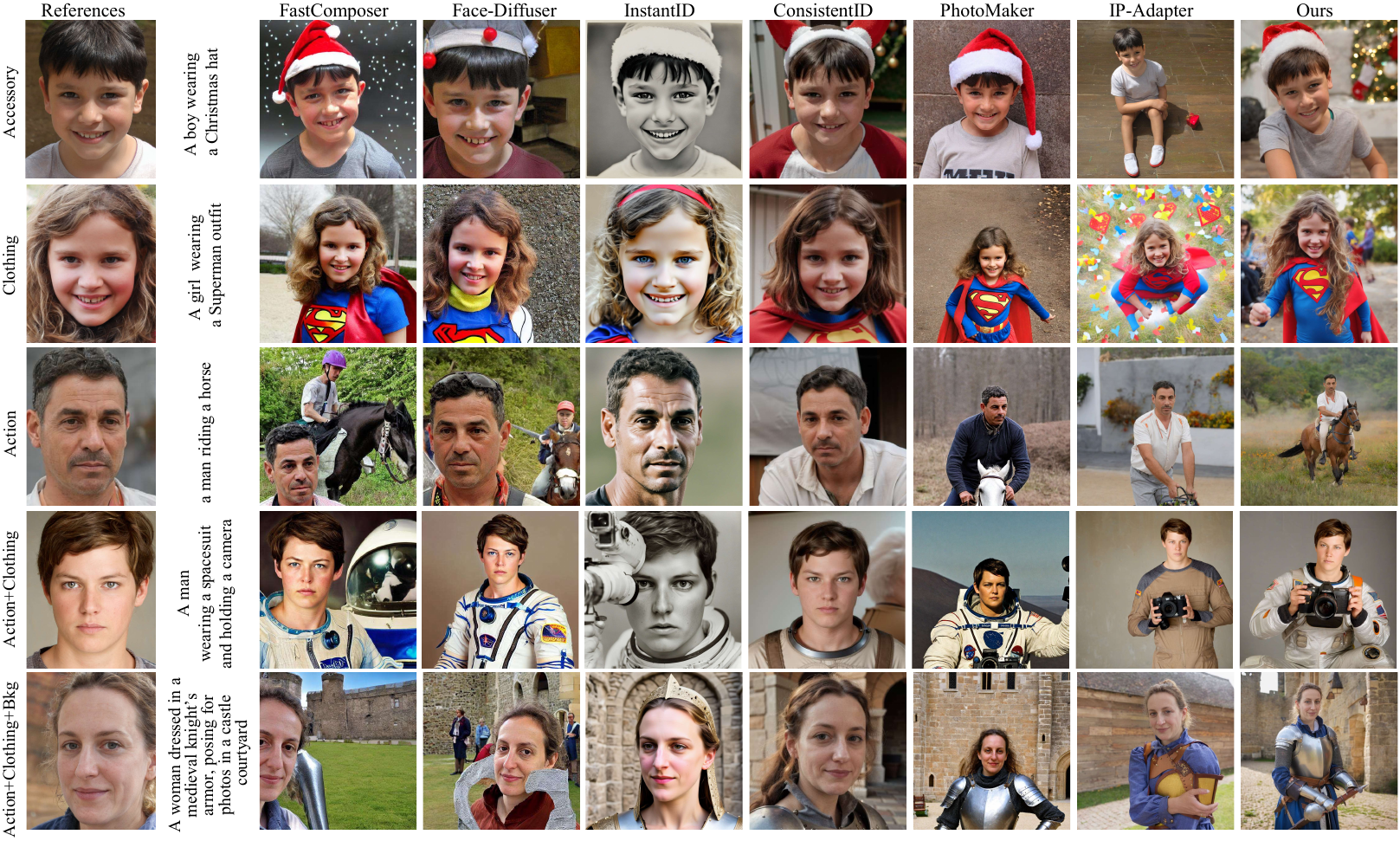}
    \caption{Qualitative comparison with several SOTA human image customized generation methods.}
    \label{fig:single_compare}
\end{figure*}

%% file: 4_experiments.tex
\section{Experiments}

\subsection{Implementation Details} 
We adopt SDXL as our base model and refer to the source code of IP-Adapter \cite{ye2023ip} for implementation and use the FFHQ dataset processed by FastComposer \cite{xiao2023fastcomposer} for training. We use cropped face images as image prompts for IEA and TCA. For each cropped face image, we apply random flipping and rotation for data augmentation. Our model is trained on a single NVIDIA A6000 GPU with a batch size of 4. The training process takes approximately 6 hours to complete. We utilize the Adam optimizer with a learning rate set to 1e-5. During the training process, only the parameters of IEA and TCA are updated. During inference, we use the DDIM sampler for 50 steps, with the guidance scale set to 5.0. The generated image size is 1024×1024.

\subsection{Experiment Setting}

\noindent{\textbf{Comparison Methods}}. We compare our approach with several SOTA methods: FastComposer \cite{xiao2023fastcomposer}, FaceDiffuser \cite{wang2024high}, IP-Adapter \cite{ye2023ip}, PhotoMaker \cite{li2024photomaker}, InstantID \cite{wang2024instantid}, ConsistentID \cite{huang2024consistentid}.

\noindent{\textbf{Evaluation Metrics}}.
To comprehensively evaluate the performance of our method, we employed two widely used metrics in related works \cite {li2024photomaker,wang2024instantid,huang2024consistentid,xiao2023fastcomposer}: the CLIP image-text similarity score (CLIP-IT) and the face similarity score (face score). These metrics assess the text consistency and facial fidelity of the generated images, respectively. Moreover, we introduced two human preference metrics, PickScore \cite{Kirstain2023PickaPicAO} and HPS \cite{wu2023better}, to assess whether the generated results align with human aesthetic preferences.

\noindent{\textbf{Test Data}}. We collected 30 individual image samples. The dataset was curated to ensure balanced gender and age distribution, thereby enhancing its representativeness and diversity. To comprehensively evaluate the performance of our method, we prepared 40 prompts that cover actions, clothing, accessories, and scenes. Additionally, we combined these prompts to generate complex text prompts for further testing. For each text prompt, we generated four images, resulting in a total of 4.8K images for evaluation. More details are listed in the  supplementary material.

\subsection{Comparison with SOTA Methods}

\noindent{\textbf{Qualitative Results}}.
Figure \ref{fig:single_compare} presents a qualitative comparison between our method and other approaches. We showcase the generated results across five categories of textual descriptions: accessory, clothing, action, action + clothing, and action + clothing + background. Although InstantID \cite{wang2024instantid} and ConsistentID \cite{huang2024consistentid} tend to generate high-fidelity portraits, they often lose consistency with textual semantics. IP-Adapter \cite{ye2023ip} improves text consistency to some extent but still loses some textual elements, especially under composite multi-category textual conditions (e.g., "riding horse" in the third row and "spacesuit" in the fourth row). PhotoMaker \cite{li2024photomaker} exhibits good text consistency, and the generated images align well with the textual prompts. However, it fails to produce high-fidelity human images. In contrast, our method not only generates high-fidelity human images but also maintains consistency with textual prompts. Additionally, our approach can generate images with complex and realistic backgrounds, further demonstrating its diverse generation capabilities.

\begin{figure}[t]
    \centering
    \includegraphics[width=0.8\linewidth]{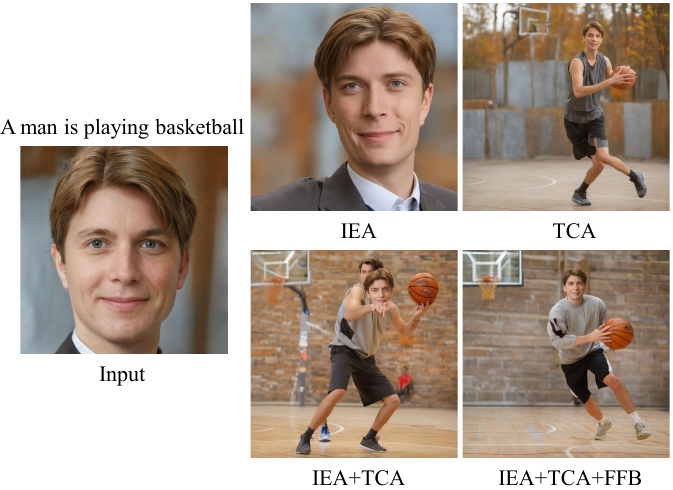} 
    \caption{Qualitative ablation study for core components of DP-Adapter.} 
    \label{fig:ablation} 
\end{figure}

\noindent{\textbf{Quantitative Results}}. 
In Table \ref{tab:compare}, we present the quantitative results of our method compared with other state-of-the-art (SOTA) methods. It is evident that these SOTA methods struggle to achieve both high-fidelity generation and text semantic consistency. For instance, PhotoMaker achieves the highest image-text similarity but has poor face fidelity, with a Face Score of 65.42. While InstantID, ConsistentID, and IP-Adapter demonstrate high Face Scores, their text semantic consistency is notably poor, with a maximum CLIP-IT score of only 22.76. FastComposer and Face-Diffuser perform poorly on both metrics. In contrast, our method achieves the highest Face Score (81.06) and the second-highest CLIP-IT score (25.07, just behind PhotoMaker), indicating that our approach successfully enhances both high fidelity and text semantic consistency. Furthermore, when evaluating alignment with human aesthetic preferences, our method outperforms all compared approaches, as evidenced by the highest PickScore (21.97) and HPS (21.31). This further demonstrates its superiority in generating images that align with human aesthetics and underscores its practical applicability.

\input{Tabs/table1}

\input{Tabs/ablation}

\subsection{Ablation Study}

Table 2 illustrates the efficacy of our method's components under different configurations. Employing the IEA alone results in high facial fidelity (79.72) yet low text consistency (19.05). The TCA, when used alone, improves text consistency (24.25) but reduces facial fidelity (64.48). Combining IEA and TCA enhances both metrics, increasing CLIP-IT to 24.81 and Face Score to 80.26. The addition of the FFB component further optimizes performance, with CLIP-IT rising to 25.07 and Face Score to 81.06. The aforementioned results confirm the effectiveness of the IEA, TCA, and FFB components.

We also present qualitative ablation results, as shown in Figure \ref{fig:ablation}. It can be observed that while IEA cannot generate content corresponding to the text prompt, it can produce high-fidelity human images, indicating that IEA focuses on improving human fidelity. In contrast, TCA can generate highly detailed scenes, such as basketball courts, but the facial details are poor. Naively blending IEA and TCA with a hard mask results in many artifacts. However, FFB, through fine-grained feature blending, effectively eliminates these artifacts, producing natural human images that maintain both high fidelity and text consistency.

\subsection{Applications}

\begin{figure}[t]
    \centering
    \includegraphics[width=\linewidth]{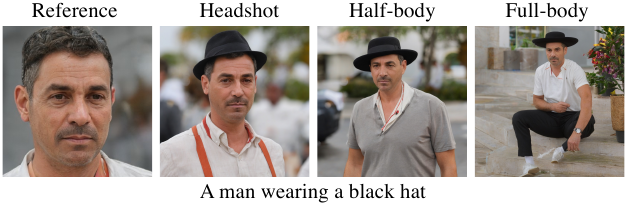} 
    \caption{Applications of controllable headshot-to-full- body portrait generation.} 
    \label{fig:p2f} 
\end{figure}

\noindent{\textbf{Controllable Headshot-to-Full-Body Portrait Generation}}.
Most current methods lack the flexibility to control the generation from headshot to full-body images. In contrast, our method allows for controllable generation of headshot, half-body, and full-body images by adjusting the $\alpha$ weight: see Figure \ref{fig:p2f}. This capability  is beyond the reach of most existing methods.

\begin{figure}[t]
    \centering
    \includegraphics[width=\linewidth]{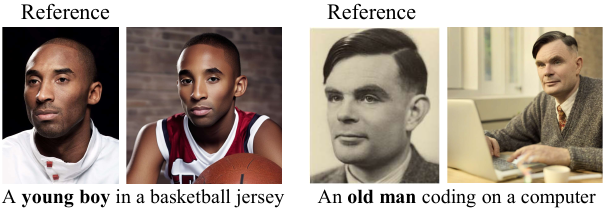} 
    \caption{Applications of changing human age.} 
    \label{fig:age} 
\end{figure}

\noindent{\textbf{Age Editing}}.
As illustrated in Figure \ref{fig:age}, our method allows for editing the age of the person in the reference image.

\begin{figure}[t]
    \centering
    \includegraphics[width=\linewidth]{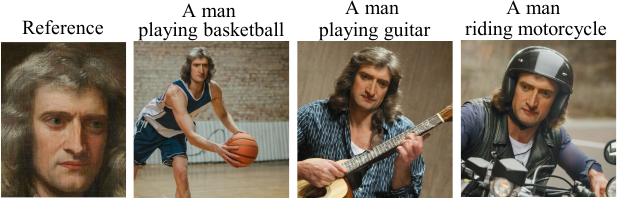} 
    \caption{Applications of bringing old photos to life.} 
    \label{fig:old2real} 
\end{figure}

\noindent{\textbf{Old-Photo to Reality}}.
Our method also supports bringing person from old photographs into real life. As shown in Figure \ref{fig:old2real}, we can generate images of Newton in various real-life scenarios.

\begin{figure}[H]
    \centering
    \includegraphics[width=\linewidth]{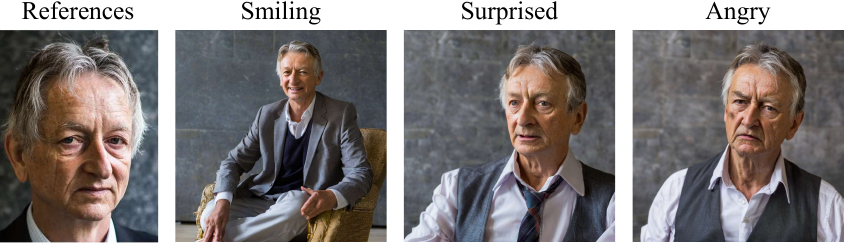} 
    \caption{Applications of expression editing.} 
    \label{fig:expression} 
\end{figure}

\noindent{\textbf{Expression Editing}}.
As shown in Figure \ref{fig:expression}, our method also supports specifying facial expressions, such as smiling, surprised, or angry, during customized generation.

\subsection{Ethics Check}
We place a high priority on ethical considerations. Like other methods \cite{ruiz2023dreambooth, li2024photomaker, wang2024instantid}, our approach faces certain ethical risks. However, its primary aim is to provide users with a respectful and meaningful personalization solution, such as creating commemorative items or enhancing interpersonal relationships. To prevent misuse of the method, we have implemented stringent security checks. We strictly prohibit any form of abuse or generation of harmful content.

%% file: Tabs/table1.tex
\begin{table}[]
\resizebox{\linewidth}{!}{
\begin{tabular}{c|cccc}
\hline
                                                           & \multicolumn{4}{c}{Quantitative Result}          \\ \hline
Method                                    & CLIP-IT        & Face Score     & PickScore   & HPS   \\
FastComposer \cite{xiao2023fastcomposer}  & 22.67          & 67.91          & 20.11       & 19.69   \\
Face-Diffuser \cite{wang2024high}         & 22.95          & 69.70          & 20.31       & 19.88   \\
PhotoMaker \cite{li2024photomaker}        & \textbf{27.36} & 65.42          & 21.36       & 21.24   \\
InstantID \cite{wang2024instantid}        & 22.64          & 79.91          & 17.51       & 20.43   \\
ConsistentID \cite{huang2024consistentid} & 21.01          & 80.78          & 20.62       & 20.23   \\
IP-Adapter \cite{ye2023ip}                & 22.76          & 80.78          & 20.76       & 20.52   \\
Ours                                      & 25.07          & \textbf{81.06} & \textbf{21.97} & \textbf{21.31} \\ \hline
\end{tabular}
}
\caption{Quantitative comparison with several SOTA human image customized generation method.}
\label{tab:compare}
\end{table}

%% file: Tabs/ablation.tex



\begin{table}[t]
\centering
\label{tab:ablation}
\resizebox{0.8 \linewidth}{!}{%
\begin{tabular}{ccc}
\hline
Settings    & CLIP-IT & Face Score \\ \hline
IEA         & 19.05   & 79.72     \\
TCA         & 24.25   & 64.48        \\
IEA+TCA     & 24.81   & 80.26      \\
IEA+TCA+FFB & 25.07   & 81.06      \\ \hline
\end{tabular}
}
\caption{Quantitative ablation study of the core components of DP-Adapter.}
\end{table}

%% file: 5_conclusion.tex
\section{Conclusion}
In this work, we introduced the Dual-Pathway Adapter (DP-Adapter), a module that improves both identity preservation and text consistency in personalized human image generation. By separating the visual and textual influences into different pathways, the DP-Adapter reduces the interference that often affects existing methods. Through extensive experiments and diverse applications, we demonstrated that DP-Adapter not only outperforms state-of-the-art methods in both quality and consistency but also provides a versatile and robust tool for personalized human image generation. The DP-Adapter faces limitations in handling corner cases where textual prompts require certain artistic styles. Future work will focus on extending the approach to general subject-driven synthesis and handling a broader range of challenging scenarios.